\documentstyle{article}
\begin{document}
\LARGE
\begin{center}
\bf Quantum Black Hole 
\vspace*{0.7in}
\normalsize \large \rm 

Wu Zhong Chao

Specola Vaticana

Vatican City State

\vspace*{0.55in}
\large
\bf
Abstract
\end{center}
\vspace*{.1in}
\rm
\normalsize
\vspace*{0.1in}

Creation of a black hole in quantum cosmology is the third way of
black hole formation. In contrast to the gravitational collapse
from a massive body in astrophysics or from the quantum
fluctuation of matter fields in the very early universe, in the
quantum cosmology scenario the black hole is essentially created
from nothing. The black hole originates from a constrained
gravitational instanton. The probability of creation for all
kinds of single black holes in the Kerr-Newman family, at the
semi-classical
level, is the
exponential of the total entropy of the universe, or one quarter
of the sum of both the black hole and the cosmological horizon
areas.
The de Sitter spacetime is the most probable evolution at the
Planckian era.

\vspace*{0.3in}

PACS number(s): 98.80.Hw, 98.80.Bp, 04.60.Kz, 04.70.Dy


\vspace*{0.5in}

e-mail: wu@axp3g9.icra.it

Permanent address: Dept. of Physics, Beijing Normal University,
Beijing, P.R.
China

\pagebreak

\rm 

\normalsize

It is well known in astrophysics that a black hole can be formed
in two ways. The first is the gravitational collapse of a
massive body. If the mass of a star exceeds about twice that of
the Sun, a black hole will be its ultimate corpse. 
The second way of formation of a so-called primordial black hole
originates from the fluctuation of matter distribution in the
early universe. In the Big Bang model, the matter content can be
classically described [1][2], while in the inflationary universe
the matter content is attributed to the quantum fluctuation of
the Higgs scalar [3]. The mass for the black hole formed in this
way is very low. 

Strictly speaking, black holes formed through the second way can
hardly be regarded as primordial. A true primordial black hole
should be created at the moment of the birth of the universe.
Therefore, we are introducing the third way, i.e. the black
hole creation in the quantum cosmology scenario. In this scenario
both spacetime and matter fields are quantized, and most
dramatically, the black hole is essentially created from nothing.

Over the last decade there have been several attempts to deal
with this problem; however, their results are not conclusive
[4][5]. Recently, many studies have been carried out for the
creation of black hole pairs [6][7][8][9][10][11]. However, the
most interesting case is the creation of a single primordial
black hole, which is the topic of this article. 

In the No-Boundary Universe a Lorentzian evolution of the
universe emanates from  Euclidean
manifolds through a quantum transition at a 3-surface $\Sigma$
with the matter field $\phi$ on it. Its probability can be
written as a path integral [12]
\begin{equation}
P = \Psi^\star \Psi = \int_{C} d [g_{\mu \nu}] d[\phi]
\exp(-\bar{I} ([g_{\mu \nu}, \phi ]),
\end{equation}
where class $C$ is all no-boundary compact Euclidean
4-metrics and matter field configurations which agree with the
given 3-metric $h_{ij}$ and matter field $\phi$ on $\Sigma$. Here
$\bar{I}$ means the Euclidean action.

The Euclidean action for the gravitational part for a smooth
spacetime
manifold $M$ with boundary $\partial M$ is
\begin{equation}
\bar{I} = - \frac{1}{16\pi} \int_M (R - 2\Lambda) -
\frac{1}{8\pi} \int_{\partial M}  K,
\end{equation}
where $\Lambda$ is the cosmological constant, $R$ is the scalar
curvature, $K$ is the trace of the second fundamental form of the
boundary.

Here, we do not restrict class $C$ to contain regular metrics
only, since the derivation of Eq. (1) from the ground state
proposal of Hartle and Hawking has already
inevitably led to some jump discontinuities in the extrinsic
curvature at $\Sigma$. To make the theory consistent, one has to
allow the discontinuity to occur anywhere in $M$.

The dominant contribution to the path integral (2) comes from
stationary action trajectories, which are the saddle
points of the path integral. The stationary action trajectories
should meet all requirements on the 3-surface $\Sigma$ and other
restrictions. At the $WKB$ level, the exponential of the negative
of the stationary action is the probability of the corresponding
Lorentzian trajectory.

In some sense, the set of all regular metrics is not complete,
since for many cases, under the usual regularity conditions and
the requirements at the equator $\Sigma$, there may not exist any
stationary action metric, i.e. a gravitational instanton which
is defined as a Euclidean solution to the Einstein field
equation. It seems reasonable to include metrics with jump
discontinuities of extrinsic curvature and with their degenerate
cases, i.e. the conical singularities, into class $C$ 
[13]. Within the extended class $C$ one can hopefully
find a stationary action trajectory, i.e. a constrained
gravitational 
instanton with some mild singularities in the absence of a
regular instanton. 
For our consideration, the singularity can only occur at some
locations 
on the given 3-metric $\Sigma$. The stationary action
trajectory satisfies the usual Einstein field equation except for
the singularities. One can rephrase this by saying that the
metric obeys
the generalized Einstein equation in the whole manifold. The
extrinsic
curvature will not vanish at the singularity locations at
$\Sigma$. Since
this result is derived from first principles and one is dealing
with the action itself, instead of the Einstein equation, in
quantum cosmology, one should not feel upset about this
situation.

In general, a wave packet of the wave function of the universe
represents an ensemble of classical trajectories. Under our
scheme, the most probable trajectory associated with an instanton
can be singled out [14]. Thus, quantum cosmology obtains its
complete power of prediction. It means there is no further degree
of
freedom except for a physical time as long as the model is
well-defined.

It is believed that the Planckian era of the universe can be
described by a de Sitter spacetime with some effective
cosmological constant $\Lambda$ [15]. Therefore, one is
interested in the
black hole creation in this background. If there is no black hole
in the universe, then one can get a regular instanton $S^4$. If
there is, then the restrictions are strong enough to allow one to
have a constrained nonregular instanton only, and the
corresponding
stationary action
will take a relatively greater value.  Therefore, the probability
of a universe with a black hole is always smaller than one
without a black hole.

First, we can consider the spherically symmetric vacuum case
[16]. The Euclidean
Schwarzschild-de Sitter metric with mass parameter $m$ is [17],
\begin{equation}
ds^2 = \left (1- \frac{2m}{r} - \frac{\Lambda r^2}{3} \right
)d\tau^2 +\left (1- \frac{2m}{r} - \frac{\Lambda r^2}{3}\right
)^{-1}dr^2 + r^2 (d\theta^2 + \sin^2 \theta d\phi^2).
\end{equation}
In general, one can use $r_2, r_3$ to denote the black hole and
cosmological 
horizons, which are the two positive roots of the expression $1-
\frac{2m}{r} - 
\frac{\Lambda r^2}{3} $ for this case. If
$0 \leq m \leq m_c = \Lambda^{-1/2}/3$, then one has an Euclidean
sector
$ r_2 \leq r \leq r_3$. For the extreme case $ m = m_c$ the
sector 
degenerates into the $S^2 \times S^2$ Nariai space. The Nariai
spacetime is identified as a pair of black holes punched through
the $S^4$ space.

In the $(\tau -  r)$ plane $r = r_2$ is an axis of symmetry and
the imaginary time coordinate $\tau$ is identified with period
$\beta_2$, whose reciprocal is the Hawking temperature. This
makes the Euclidean manifold regular at the black hole horizon.
One can also apply this procedure to the cosmological horizon
with period $\beta_3$, whose reciprocal is the Gibbons-Hawking
temperature. For the $S^2 \times S^2$ case these two horizons are
identical, thus one obtains a regular instanton. Except for the
$S^2 \times S^2$ spacetime, one cannot simultaneously regularize
it at both horizons because of the
inequality $\beta_2^{-1} > \beta_3^{-1}$.

Now we are going to construct a constrained
gravitational instanton. One can have two cuts at $\tau =
consts.$ between $r = r_2$ and $r = r_3$. Then the $f_2$-fold
cover turns the $(\tau - r)$ plane into a cone with a deficit
angle $2\pi (1-f_2)$ at the black hole horizon. In a similar way
one can have an $f_3$-fold cover at the cosmological horizon.
Both $f_2$ and $f_3$ can take any pair of real numbers with the
relation 
\begin{equation}
f_2 \beta_2 = f_3 \beta_3.
\end{equation}
This manifold satisfies the usual Einstein equation except for
the conical singularities at at least one of the two horizons.
The variation
calculation of the action requires that the manifold obeys the
Einstein
equation everywhere with the possible exception at the transition
surface where the constraints are imposed. We assume the quantum
tunneling will occur at the equator which are two $\tau = const.$
sections, say $\tau = \pm f_2\beta_2/4$, passing through the two
horizon. Therefore to check whether we
have obtained a constrained instanton, it is only necessary to
prove
that the action is stationary with respect to the parameter $f_2$
or $f_3$, i.e, the only degree of freedom left. 

If $f_2$ or $f_3$ is different from $1$, then the cone at the
black hole or cosmological horizon will have an extra
contribution to the action of the manifold. 
Since the integral of $K$ with respect to the $3$-area in the
boundary term of the action (2) is the area increase rate along
its normal, then the extra contribution due to the conical
singularities can be considered as the degenerate form shown
below
\begin{equation}
\bar{I}_{i,cone} = - \frac{1}{8 \pi}\cdot 4\pi r_i^2\cdot 2\pi
(1 - f_i).\;\;\; (i = 2, 3)
\end{equation}
Thus, the total action 
can be calculated
\begin{equation}
\bar{I}_{total} = -\frac{f_2 \beta_2\Lambda}{6} (r_3^3 -
r_2^3) + \sum_{i = 2,3}\bar{I}_{i,cone},
\end{equation}
where the first term of the right hand side is due to the volume
contribution.

Substituting Eqs. (4) and (5) into Eq. (6), one can obtain 
\begin{equation}
\bar{I}_{total} = - \pi (r^2_2 + r^2_3).
\end{equation}
This is one quarter of the negative of the sum of these two
horizon areas. One quarter of the sum is the total entropy of the
universe. 

It is remarkable to note that the action is independent of the
choice of $f_2$ or $f_3$. This means that our constructed
manifold has a stationary action and is qualified as a
constrained gravitational instanton. Therefore it can be used
for
the $WKB$ approximation to the wave function.  This phenomenon
also occurs for the whole family of Kerr-Newman black holes as we
shall discuss below. Nature has a great propensity for black
holes!
Therefore, the
creation probability of a Schwarzschild black hole in the de
Sitter background, at the $WKB$ level, is
\begin{equation}
P_m \approx \exp (\pi(r^2_2 + r^2_3)).
\end{equation} 

Our result implies that no
matter which flat fragment of the manifold is chosen, the same
black hole should be created with
the same probability. Of course, the most dramatic case is that
of no volume, i.e. $f_2 = f_3 = 0$.

Formula (8) interposes the following two extreme cases [9].
First for the de Sitter case with $m = 0$, $P_{0} \approx \exp
(3\pi \Lambda^{-1})$,
and second for the Nariai case with $m = m_c$, $P_{m_c} \approx
\exp (2\pi \Lambda^{-1})$.

The Schwarzschild black hole case is the simplest, since one can
easily glue the north portion and the south portion of the
instanton at the equator $\Sigma$. One can also consider the
Reissner-Nordstr$\rm\ddot{o}$m black hole case[16]. When the
black hole is magnetically
charged, then the matter field is represented by the vector
potential $A= Q(1- \cos \theta )d \phi$ over the $S^2$
space, ($\phi - \theta$) sector, where $Q$ is the charge. There
is no obstacle to gluing,
since the magnetic field is continuous at $\Sigma$. For the
electrically charged case, one can choose the vector potential $A
= -iQr^{-2} \tau dr$. Since the vector potential $A$ at $\Sigma$
does not uniquely define the electric
field there, then the formula (1) does not represent the
probability of a black hole with an electric charge. In fact, the
configuration of the wave function is the three geometry and the
momentum $\omega$, which is canonically conjugate to the electric
charge $Q$ and is defined by the integral of $A$ around $S^1$
sector of $\Sigma$. Then one can get the wave function
$\Psi(Q, h_{ij})$ for the given electric charge through the
Fourier transformation [10][11]
\begin{equation}
\Psi (Q, h_{ij}) = \frac{1}{2\pi}\int^\infty_{-\infty} d\omega
e^{i\omega Q} \Psi (\omega, h_{ij}).
\end{equation}
At the $WKB$ level, the transformation is equivalent to adding
the
following extra term 
\begin{equation}
f_2 \beta_2 Q^2 (r^{-1}_2 - r^{-1}_3)
\end{equation}
to the action 
\begin{equation}
\bar{I} = -\frac{f_2 \beta_2 \Lambda}{6} (r^3_3 -
r^3_2) \pm
\frac{f_2 \beta_2 Q^2}{2}(r^{-1}_2 - r^{-1}_3) +
\sum_{i=2,3}\bar{I}_{i,cone},
\end{equation}
where $+$ is for the magnetic case and $-$ is for the electric
case. Therefore, the Fourier transformation  will iron out the
sign 
difference of the action
terms due to magnetic and electric fields and thus recover the
duality between magnetically and electrically charged black
holes. The total action is stationary again. The probability for
Reissner-Nordstr$\rm\ddot{o}$m black hole creation is also
expressed by formula (8). All known results on the probability of
black hole creation [6][7][8][9][10][11] can be
derived as special cases from this quite universal formula.

A similar consideration should be made for the rotation of a
black
hole [16]. Again, the 3-geometry of $\Sigma$  determines the
angular differentiation between the black hole and cosmological
horizons instead of the angular momentum of the hole. One has
to use another Fourier transformation relating them to obtain
the wave function for a given angular momentum. The probability
of Kerr-Newman black hole creation is also the exponential of one
quarter of the sum of the black hole and cosmological
horizon areas, or the exponential of the total entropy of the
universe.

Our calculation
has very clearly shown that the gravitational entropy is
associated with the spacetime topology. 
From the no-hair theorem, a stationary black hole in the de
Sitter spacetime background is characterized by three
parameters only, mass, charge and angular momentum, so the
problem of quantum creation of a single black hole at the birth
of the universe is completely resolved.

It can be shown that the probability is an exponentially
decreasing function of the mass, magnitude of charge and angular
momentum of the black hole. Therefore, the de Sitter spacetime is
the most probable evolution of the universe at the Planckian era.

\vspace*{0.1in}

\bf Acknowledgment:

\vspace*{0.1in}
\rm

I would like to thank G. Coyne of Specola Vaticana and
S.W. Hawking of Cambridge University for their
hospitality. 
\vspace*{0.1in}

\bf References:

\vspace*{0.1in}
\rm

1. S.W. Hawking, \it Mon. Not. R. Astro. Soc. \rm
\underline{152}, 75 (1971).

2. B. Carr and S.W. Hawking, \it Mon. Not. R. Astro. Soc. \rm
\underline{168}, 399 (1974).

3. Z.C. Wu, \it Phys. Rev. \bf D\rm\underline{30}, 286 (1984). 

4. L.Z. Fang and M. Li, \it Phys. Lett. \rm \bf B\rm
\underline{169}, 28 (1986).

5. Z.C. Wu, in \it Proceeding of the Fourth Marcel Grossman
Meeting \rm  (North Holland, 1986).

6. P. Ginsparg and M.J. Perry, \it Nucl. Phys. \rm \bf
B\rm\underline{222}, 245 (1983).

7. F. Mellor and I. Moss,   \it Phys. Lett. \rm \bf B\rm 
\underline{222}, 361 (1989); \it Class. Quantum Grav. \rm \bf
\underline{6}\rm, 1379 (1989). 

8. I.J. Romans, \it Nucl. Phys. \rm \bf B\rm \underline{383},
395 (1992).

9. R. Bousso and S.W. Hawking, \it Phys. Rev. \rm \bf D\rm 
\underline{52}, 5659 (1995).

10. S.W. Hawking and S.F. Ross, \it Phys. Rev. \rm \bf
D\rm\underline{52}, 5865
 (1995).

11. R.B. Mann and S.F. Ross, \it Phys. Rev. \rm \bf
D\rm\underline{52}, 2254 (1995).

12. J.B. Hartle and S.W. Hawking, \it Phys. Rev. \rm \bf D\rm
\underline{28}, 2960 (1983).

13. G. Hayward and J. Louko, \it Phys. Rev. \rm \bf D\rm
\underline{42}, 4032 (1990).

14. X.M. Hu and Z.C. Wu, \it Phys. Lett. \rm \bf B\rm
\underline{149}, 87 (1984).

15. S.W. Hawking, \it Nucl. Phys. \rm \bf B\rm \underline{239},
257 (1984).

16. Z.C. Wu, \it Int. J. Mod. Phys. \rm \bf D\rm \underline{6},
199 (1997); \it Prog. Theo. Phys. \rm \underline{97}, 859 (1997);
\it Prog. Theo. Phys. \rm \underline{97}, 873 (1997).

17. G.W. Gibbons and S.W. Hawking, \it Phys. Rev. \bf D\rm
\underline{15}, 2738 (1977).

\end{document}